# The detection of professional fraud in automobile insurance using social network analysis

**Arezo Bodaghi, MD,**

School of Industrial and Systems Engineering,

Tarbiat Modares University, Iran

**Babak Teimourpour, Ph.D.,**

**Assistant Professor**

School of Industrial and Systems Engineering,

Tarbiat Modares University, Iran

**Abstract**

The Automobile Insurance Fraud is one of the main challenges for insurance companies. This form of fraud is performed either opportunistic or professional occurring through group cooperation that leads to greater financial losses, while most presented methods thus far are unsuited for flagging these groups. The article has put forward a new approach for identification, representation, and analysis of organized fraudulent groups in automobile insurance through focusing on structural aspects of networks, and cycles in particular, that demonstrate the occurrence of potential fraud. Suspicious groups have been detected by applying cycle detection algorithms (using both DFS, BFS trees), afterward, the probability of being fraudulent for suspicious components were investigated to reveal fraudulent groups with the maximum likelihood, and their reviews were prioritized. The actual data of Iran Insurance Company is used for evaluating the provided approach. As a result, the detection of cycles is not only more efficient, accurate, but also less time-consuming in comparison with previous methods for finding such groups.

## 1- Introduction

Fraud is committed in various fields such as insurance (Ormerod et al. 2010; Li et al. 2008; Atwood et al. 2006), credit card (Weston et al. 2008; Dal Pozzolo et al. 2014), telecommunications (Estevez, 2006), and financial communications (Kirkos et al. 2007; Kotsiantis et al. 2006; Holton, 2009). Insurance fraud is one of the most frequent types of fraud to undertake. This type of fraud can take place in many forms with the simple objective of gaining money (Almedia, 2009).

One of these domains is car insurance in which fraudsters (policyholders) setting planned traffic accidents up and file fake insurance claims (e.g. inflating costs) to obtain an illicit benefit from their insurance policy (Ayuso et al. 2011). It has been reported that Almost 21% to 36% of auto-insurance claims contain elements of suspected fraud but only less than 3% of them are prosecuted (Nian et al. 2016).

There are two different types of fraud, including opportunistic, and professional fraud that the second type is committed by organized groups. Although the organized fraud is perpetrated fewer than the opportunistic insurance fraud, the majority of revenue outflow (financial losses) is due to these groups (a White paper, 2012). According to Bolton and Han (2002), fraud detection would be difficult due to many reasons. The first one is, involving high volume of data, which are constantly evolving. In reality, for processing these sets of data, the fast, the novel, and efficient algorithms are entailed. Moreover, in terms of cost, it is evident that undertaking a detailed analysis of all records is too much expensive. Here the issues of effectiveness enter; indeed, many legitimate records exist for every person that an effective method should detect fraudulent records correctly.

Traditional systems for fraud detection are only able to find fraudulent customers (opportunistic fraud), whereas more professional fraudsters will be overlooked (a white paper, Roberts, 2010). In other words, opportunistic fraud is a continuous issue for insurers, whereas the more remarkable challenge comes from professional fraud, and such organized groups of perpetrators impose the greatest cost upon insurers. Fraudulent groups are being arranged by fraudsters in order to employ different individuals for doing some works, and using the newest technologies to be at least one step in front of insurers. They know properly that insurers and law enforcements officials utilize what kind of tools, and information (Smallwood and Breading, 2011). Due to aforementioned reasons, it is imperative for insurance companies to consider relevant methods for finding organized fraud groups, and promulgating them in the future.

The detection of insurance fraud has been seriously taken into account in recent years. Although this issue is seen more in practical and functional fields, it is considered in terms of the academic aspects due to its negative effect on insurance pricing and on efficiency of insurance industry.

There is a wide range of data mining and machine learning techniques applied for fraud detection (all areas), opportunistic type, in particular, namely fuzzy logic, neural networks, support vector machines, logic, genetic, logistic regression, decision trees, naive Bayes, outlier detection, sampling, K-nearest neighbor, Evolutionary algorithms (Brockett et al. 2002; Pathak et al. 2005; Viaene et al. 2005; Perez et al. 2005; Viaene et al. 2002; Lim et al. 2000; Viaene et al. 2004; Hodge and Austin, 2004; Domingos et al. 2002; Caudill et al. 2005; Crocker and Tennyson, 2002; Viaene et al. 2007; Dobson et al. 2010; Artís et al. 2002; Bermúdez et al. 2008; Xu et al. 2011; Bhowmik, 2011; Brockett et al. 2006; Sundarkumar and Ravi, 2015; Kose et al. 2015). Gepp et al's research (2012) presents a comparative analysis of anticipatory performance of some data mining methods in which Decision Tree (LA) and Survival analysis (SA) were illustrated more appropriate to detect the automotive insurance fraud in comparison with Logit Analysis (LA), and Discriminant analysis (DA). Based on (Derrig, 2002), all the above-mentioned techniques have approximately same performances and none of them have excellence in others.

Despite the extensive utilization of data mining algorithms for sorting fraud out, according to (Phua et al. 2005), there are some complexities with regard to nature of data mining techniques that illustrate they might be inefficient in flagging fraudulent activities in future; firstly, a volume of data will fluctuate over time (doubtless that the volume of data will have boomed by near future). Secondly, the forms and styles of fraud are changing regularly. The third criticism is regarding introducing new patterns of suspicious activities during the near future like professional fraud that will have been generated. In the last decade, social networks play a prominent role in researches on detecting and deterring fraud in various areas such as credit card, Social online, financial trade, Internet Auction, health insurance, etc. (Eberle et al. 2010; Bindu and Thilagam, 2016; Yu et al. 2015; Chau et al. 2006; Akoglu et al. 2013; Chiu et al. 2014; Vlasselaer et al. 2015; Flynn, 2016).

To augment (strengthen) our research, we drew upon (exploited) prior studies in which social network approaches have been deployed for revealing, and obviating insurance fraud, professional type in particular.

Viaene and Dedene (2004) used the Naive Bayes method to detect fraud in auto insurance data. In this paper, they used the reinforced algorithm and compared the results with output obtained from applying the algorithm on a Naive Bayes without use of reinforced algorithms, and it was proven that the use of reinforced algorithms led to more precise consequences.

Almeida (2009) proposed the application of DFS algorithm for establishing the local social network location of each entity (node) in order to overcome posed issues by unbalanced data sets. Furthermore, he tracked general pattern down with the assistance of Gspan algorithm.

Subelj (2011) suggests an expert system which focuses on a relationship between the perpetrators of a car accident in order to prevent the automobile insurance fraud. The detection of suspicious components starts in this expert system after creation of network. Afterward, the system determines the suspicious entities in each of the detected suspicious components by displaying the social network of these entities. Unlike other solutions, this system uses network and graph to display data. In this paper, the researchers have provided an algorithm called Iterative Assessment Algorithm (IAA) to identify the fraud entities. Along with the original and inherent characteristics of entities, this algorithm explores the relationships between entities. According to research results, the proper data display is fundamentally important to detect the automobile insurance fraud.

Didimo et al. (2011) proposed a new system called VISFAN for visual analysis of financial activity networks. Using the effective tools, the VISFAN System helps the financial professionals and analysts to detect the financial crimes such as the money laundering and fraud on the network. Compared to other existing systems and methods for analysis of criminal networks, this system has some innovations.

De Zoete et al. (2015) have found how the use of a Bayesian Network can interpret the existing evidence in some linked crimes; in fact, they indicate that how this method can show the similar dependencies and links between crimes in order to identify the key people. Kose et al. (2015) studied, implemented and evaluated a new framework for detecting the fraudulent cases involved in these claims, and developed a structure for introducing new types of fraud. Moreover, used the well-known methods such as AHP and EM unsupervised ranking to detect abnormalities and increase the accuracy of the framework.

To sum up, although prevalent fraud detection methods, along with aforementioned weaknesses, are capable of cope with opportunistic fraudulent activities, and the least number of them have focused on organized collaboration of perpetrators (sophisticated fraud), our work is completely different. We mainly took professional fraud into consideration, we proposed a novel approach for detection and subsequent investigation of automobile insurance fraud (suspicious consumer claims). ). The argument has focused on groups of fraudsters (organized fraud) that their structural relations (links) in a network of collisions are in form of the cycle. In other words, we believe that for detecting professional fraud is better to find cycles (rings) rather than communities (densely connected subsets of nodes, because the cycles' entities are more likely to be fraudulent. To verify the validity of the proposed method, we used the actual data of an insurance company. At the first step, the network of collisions (the relations between two colliding participants) was constructed, then through two different algorithms, suspicious components (cycles) were detected, while they all were not necessarily dubious. In third step, the suspicious groups have been investigated through some indicators for measuring the likelihood of fraudulent activities to unveil criminal entities, and prioritizing groups review, afterward; they were visualized. We decided to take structural properties of networks into account because there are various reasons why considering these characteristics of the network is more beneficial. The foremost reason is reducing the time and volume of calculations; and improving the accuracy is the other reason. In addition, it should be applicable to larger networks with millions nodes and edges. To reach this goal, we went to the depth of the network of accidents and distinguished suspicious cycles with the help of using depth first search and breadth first search algorithms. No structural analysis of networks have been done up to now, thus, our work is completely new.

Besides, a labeled dataset is not required, and also it is effective and applicable for both small and large networks with the least consumption of time and cost, and decreasing variables' dimensionality. In different words, notwithstanding the efficiency of developed algorithms, they all are still time-consuming brought into play on large-scale networks.

In section 2 of this paper, the network concepts and its analysis will be reviewed. Next, the Methodology of proposed approach will be described in section 3; followed by the evaluation of method using analysis of real data in section 4; finally, the conclusion and suggestions regarding future research will be presented in section 5.

## 2- Social networks

Network and a mathematical graph are on the same side, and utilize an identical set of techniques for analysis of a wide range of cases; from international trade to co authorship to protein interactions (Borgatti et al. 2009). A graph is a pair G = (VG, EG) constituting vertices (nodes) and arcs (edges) where:

$$VG = \{v1, v2, \ldots, vn\}$$

$$EG = \{e1, e2, \ldots, en\} \sim \{(vi, vj) \mid vi, vj \in VG \wedge i \neq j\}$$

The V is the ensemble of nodes, and the E is a group of edges. Besides, an edge goes with two vertices, and always the intersection of V and E is assumed empty (V ∩ E = Ø) due to the prevention of ambiguity. A degree of a vertex is the number of edges at v (Diestel, 2010). $d(v)$ or deg v is accounted for the degree of each vertex so that $d(v) = |E(v)|$.

Undirected graph is a graph with edges that have no orientation (figure 1(a)), whilst a directed graph (digraph) is a graph in which a bunch of vertices are connected together through edges having directions (figure 1(b)). Multi graph is the other type of graph that has multi edges _when two same vertices are connected to each other by two or more edges. These edges are considered as multi edges (figure 1(c)) (Trudeau, 1993). Other terminologies in graph theory are path, tree, and cycle that will be explained respectively. To begin down the path that is a sequence (finite or infinite) of edges by which an array of vertices is joined up with together. Note that all vertices in a path should be unique (exclusive of possibly the first and last). Figure 1(c) is a clear example for demonstration of different paths between v1 and v3:

v1, e1, v2, e2, v3

v1, e10, v5, e6, v2, e2, v3

v1, e1, v2, e5, v5, e7, v4, e3, v3

In a finite group, the distance d (v, u) between two vertices v and u is the minimum length of the paths that connect them (Bondy and Murty, 1976). Afterward, a cycle is the path from a vertex to itself. In variant words, a cycle consists of a sequence of ($e_j \in E$, $0<j\leq n$, $vi \in V$, $0\leq i\leq n$), ($v_0$, $e_1$, $v_1$,..., $e_n$, $v_n$) where $V_0 = V_n$; when all vertices and edges are unique except for these two vertices in the sequence, the cycle is defined as a simple cycle (West, 2001). In figure 1(a), figure 1(b) two examples for cycle in an undirected and directed graph have been provided respectively.

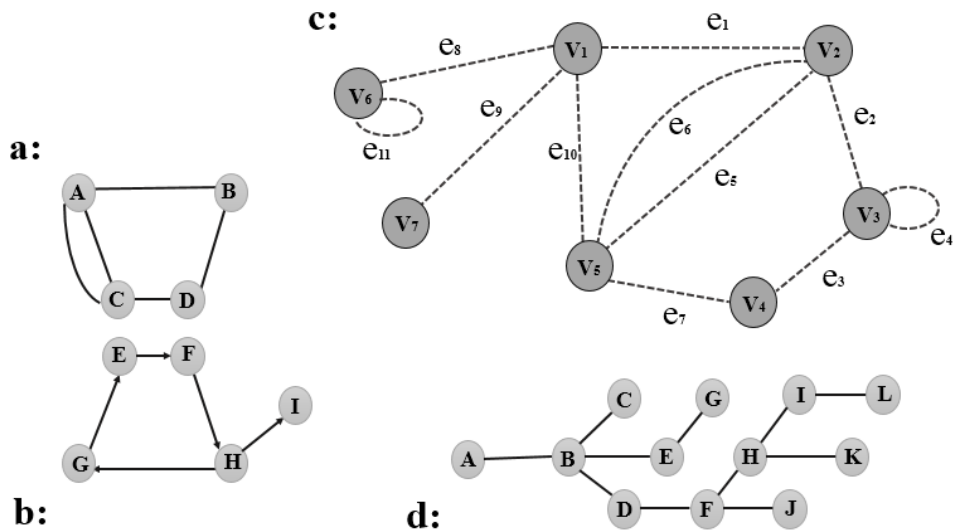

**Figure 1(a).** Undirected graph; (b) Directed graph; (c) Multi graph with multi edges; (d): Tree with only a path between two nodes

The Cycles play very important roles in graphs and have numerous applications in various fields such as the electronic network analysis, analysis of biological and chemical routes, the periodic planning, and graph drawing, and so on (Kavitha et al. 2009). All graphs are divided into two categories: cyclic and acyclic; the first type (cyclic) involving a cycle without any repeated vertices, while the second type (acyclic) has no cycles. The last one is a tree. A tree is an acyclic connected graph in which precisely one path exists connecting any two vertices (see figure 1(d)) (Gross et al. 2013).

From 20th century so far, graphs have been extensively applied for representing a large variety of systems in diverse areas (Fortunato, 2010). One of the graph theory's applications is to provide joint methods for myriad different-appearance issues. The basis of graph theory is visualization. On the one hand, the models of graph theory have a connection with real world using graphical terms. On the other hand, computational methods yielded by mathematics.

As a result, due to these interplays, graph theory has grabbed attentions (Ruohonen, 2013). Graph theory's terms and concepts could be used for representing and analysis of networks (Freeman, 1978). Network is a collection of nodes, which are connected together in pairs by edges.

Social networks are networks in which people or sometimes groups of people are considered as the nodes, and some form of their social interaction such as friendship represented in edges (Newman, 2010). The two-mode network (of cars, drivers, and collisions) of collisions will be displayed in figure 2(a), and the one-mode network (of drivers) can be seen in figure 2(b).

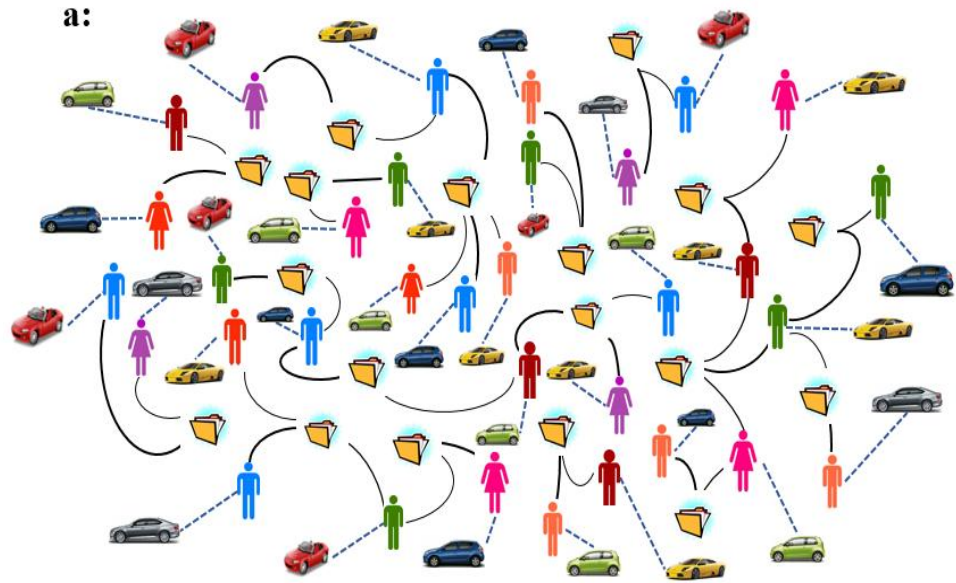

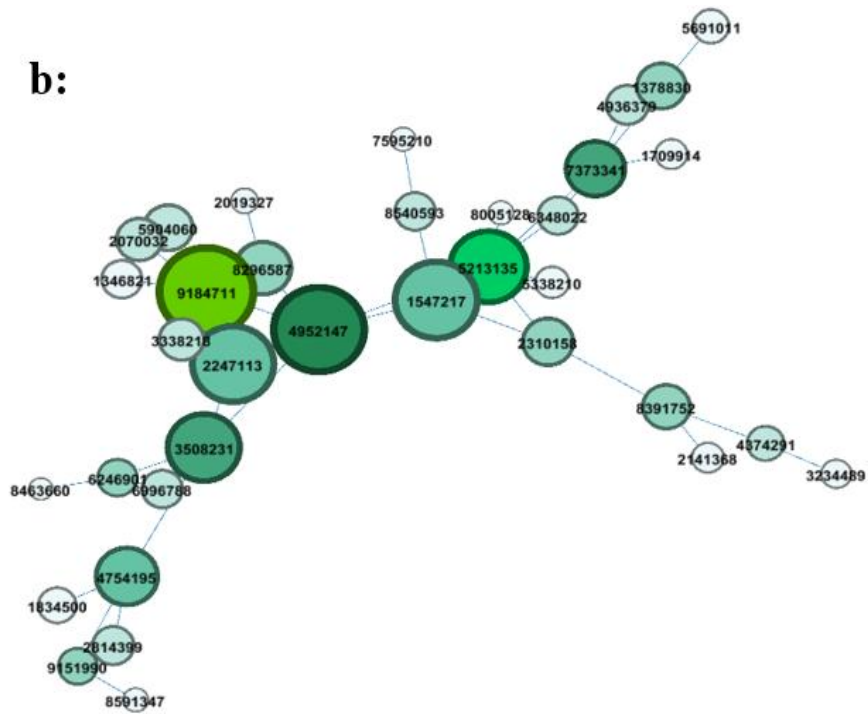

**Figure 2(a).** Network of drivers, cars, and accidents; (b) Network of drivers

There are various methods for detecting fraudulent groups within a network, including detection of communities, connected components, and Bi-connected components in a network (Pinheiro, 2011). One of the properties of networks is community structure in which the nodes are tightly (closely) connected to each other, while the connections between groups (communities) are loose. There are various types of communities in different networks, comprising papers related in a unique topic in the citation network, cycles in the network of metabolism (Girvan and Newman, 2002). Community structure conveys some essential information to a proper understanding. The identification of communities is therefore, a very important part of modern network analysis (Fortunato, 2010).

### 3- Proposed approach

As it alluded, in this study, we have focused mainly on detection of organized fraudsters through social networks, particularly structural aspects of networks. The framework of the proposed system can be seen in figure 3.

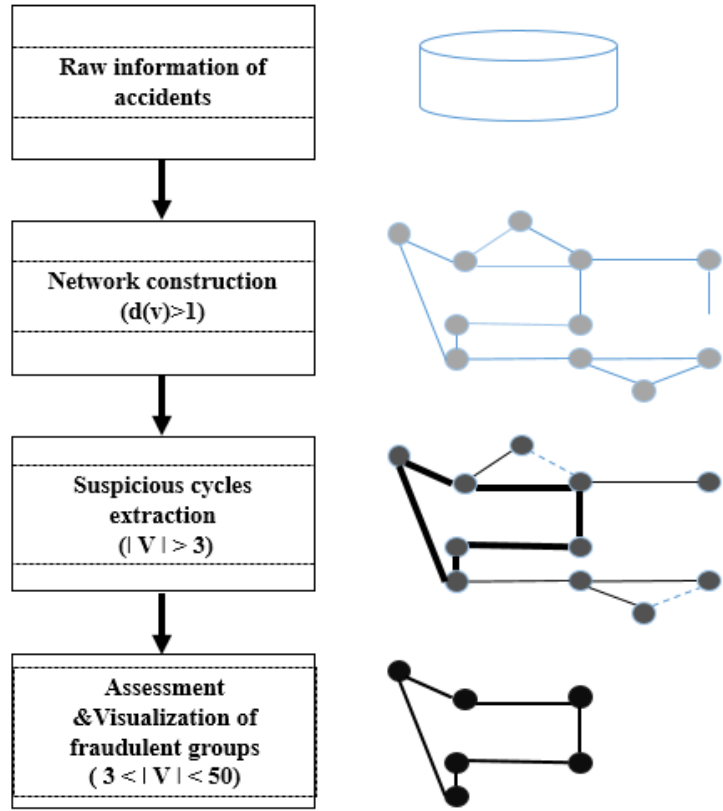

**Figure 3.** The framework of the system for detection of organized fraud

The first module was constructing the network of collisions from given dataset in which vertices correlate with drivers involved in the same collision, and edges correspond to relations (attending at same accident) between them. The next module is the investigation of the collided network structurally in order to find fraudulent collaborations. In this step, we try to find groups of drivers whose relations in the network are as the especial form (cycle). The third module is looking into detected suspicious components (cycles) and assessing the likelihood of fraud for diagnosis of fraudulent entities. In various words, the probability of occurrence of cycles with the number of nodes more than three (n>3) in the network is not accidentally, and undoubtedly it has happened with previous planning. However, the probability of cycles in which the number of nodes are high- for instance, n> 50- are close to zero, and they have accidentally been taken place. Ultimately, we are going to find cycles that the likelihoods of them being fraudulent are much more, compared to others. The explanation of each module will be seen in future sections, 4-1, 4-2, 4-3 respectively.

### *4-1 building the network of collisions*

The main purpose of the first module is to model the relationships between drivers (cars) by which collaborations have been considered naturally and comprehensively. We only have targeted main characters of an accident, including drivers and cars, and no participants or organizations, police officers, lawyers, and chiropractors because the accessibility to this information not as much easy as possible for everyone, so several possible types of network, though, being taken into account for representing the collisions, all of them are not applicable. Take the following types into consideration (see figure 4):

1). Directed – drivers / vehicles

2). Undirected – drivers/vehicles

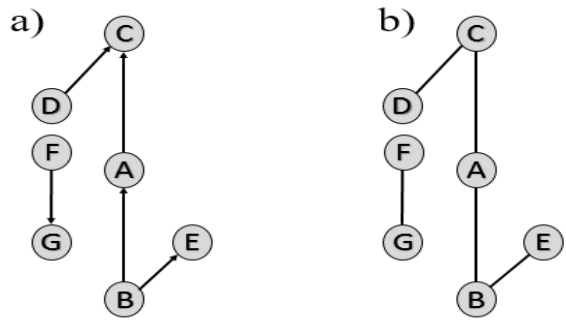

**Figure 4.** Two different types of networks: (a) the directed-drivers network, (b) the undirected-drivers network. Nodes correspond to drivers/cars, edges depict taking part in collisions.

Several guidelines have been put forward by (Šubelj et al. 2011) as to whether building networks that could be considered for more information. If there are lacked of drivers' information, network of cars could be substituted. In other words, the data on collisions –the data of drivers or cars- might have not become registered properly, in these cases; investigation of both networks could ease the shortage. Due to the lack of information on numerous individuals (drivers), we presumably have nodes with zero degrees (isolated vertices). Additionally, there are some nodes with degree one that indicate the drivers who have only an accident. Doubtless, these nodes do not contain useful information; consequently, all of them should be removed in the network of collisions and two undirected and directed network of collisions in which nodes corresponded to drivers/cars, and edges are accounted for their involvement in the same collision.

*4-2 extraction of components in form of cycles (n>3)*

In this module, one of the major challenges is the high volume of data in recent years and its increasing trend that leads to a large scale network; thus, for identification of suspicious components, the need for higher operating speed and less memory consumption is one of the main criteria for selecting proper approaches. As mentioned before, different methods could be used for selecting groups of perpetrators in a network, constituting defining a set of indicators for choosing connected components with high probability of fraud, and using RIDITs (PRIDIT) for detecting principal component analysis represented by (Bross, 1958), community detection using fast greedy (Clauset et al. 2004), walk trap, Multilevel community; however, they all are not completely suitable for detecting groups. You can see the weaknesses of mentioned approaches in comparison with our proposal in section 4.

As a result, powerful and fast algorithms are required for detecting groups in directed, and undirected networks of drivers/cars. To tackle these weaknesses we considered the cycle form components. For detection of these components, no thresholds or a particular collection of indicators is necessitated, and taking a large number of entities into account, whereas most of them not necessarily being fraudulent. The cycle bases emerge in myriad engineering applications, including the analysis of electrical circuits, figuring periodic scheduling problems out in traffic planning, and a graph drawing method (Kavitha, 2009).

There are many algorithms for discovering cycles in networks. A self-stabilizing algorithm (AFC I) was introduced by (Chaudhuri, 1999) for discovering indispensable cycles in connected undirected networks. The proposed algorithm resists toughly against short-lived faults, and initialization is not necessitated. At the end of the process, the algorithm will provide the exact number of fundamental cycles, plus a unique identifier for each of them. While the depth-first search spanning tree of the graph is known, the required time is $O(n^2)$. If not, it entails $O(n^3)$ time (n is the number of nodes).

Notwithstanding all its advantages, the uncovering of cycles might not be progressed before the construction of DFS trees. On the other hand, Bielak and Pa´nczyk, (2013) presented ASFC II algorithm for alteration of Chaudhuri's algorithm. They presented an algorithm with only 4n moves. In 2009, Lingas and Lundell presented an algorithm for revealing the shortest cycle in undirected graphs. According to this algorithm, if presence of adjacency matrix k of graph G with n nodes, a cycle with the length of 2k for even k, and 2k+2 for odd k with high likelihood will be disclosed in time $O(n^{3/2}\sqrt{\log n})$ Floyd, Brent, Nivash, and Gosper are the name of some well- known algorithms for digging cycles up in networks. For more information refer to (Nesterenko, 2012; Fisler et al. 2001; Hoàng et al. 2013). 1n 1987, Itai and Rodeh presented detection of the shortest cycle could be done in O (nm) time – n and m are the number of nodes and edges respectively- when the breadth first search (BFS) algorithm used from each node.

Hence, according to studies conducted, for detecting cycles, at first step, trees (the shortest path from each node to others) for each node, then, induced sub-graphs are entailed for all nodes of each tree in order to be compared with trees that the differences between these two graphs (tree and induced sub-graph for a same group of nodes) result in the identification of cycles. Due to the volume of network, two different algorithm have been implemented to detect cycles. And also, the result will be visualized at the end.

For detection of trees, we use BFS and DFS algorithm, finally the results were compared to each other. The Breadth First Search (BFS) is one of the graph search algorithms. The Breadth First Search strategy for graph search performs the breadth-to-breadth graph search. The algorithm starts from the root (a desired vertex is selected as the root in graphs or trees without the root) and is put them at the next first breadth. Afterward, all last breadth heads neighbors, which are not still seen, are visited and put in the next breadths. This process stops when all neighbors of vertices at the last breadth have been already seen (Cormen et al. 2002).

The Depth First Search (DFS) algorithm, from a practical perspective, is implemented according to the queue, so that the root is first put in the queue, and then the element at the beginning of the queue is pulled put, and their neighbors are checked and any neighbor, which has not been seen yet, is added to the end of a queue. This algorithm is applied to detect the spanning tree of each node in a graph; and it starts from the root (a desired vertex is selected as the root in graphs or trees without the roots), and then it investigates the neighbors of current vertex through the output edges of current vertex respectively, and finally it is implemented as the current vertex as soon as it is faced with a neighbor, with which it has not faced, through rollback method. In the case that all neighbors were not seen previously, the algorithm will be rolled back, and the implementation of the algorithm continues for vertex from which we have reached the current vertex. In other words, the algorithm goes to the depth back as far as possible, and rolled while facing with the deadlock. This process continues until all vertices reachable from the root are seen (Cormen et al. 2002). Figure 5 shows images of these two algorithms.

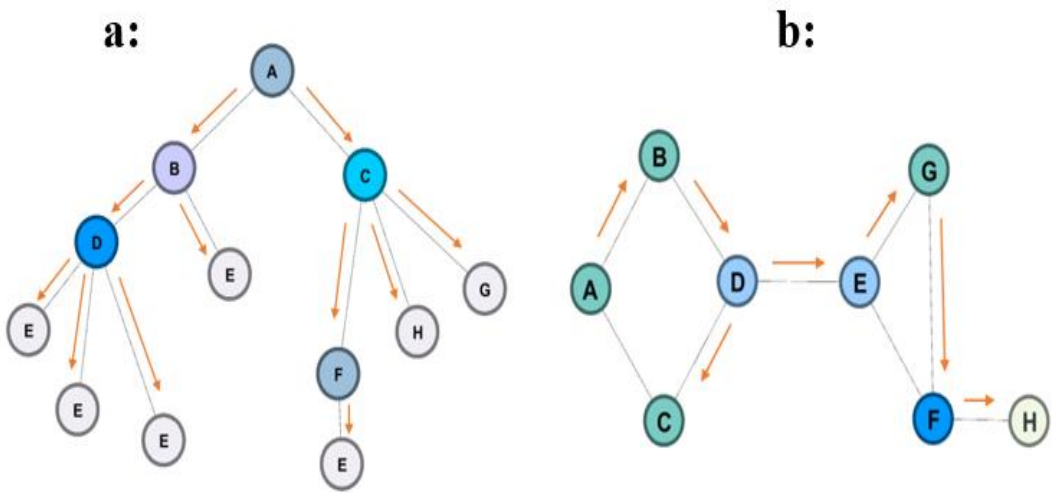

**Figure 5.** (a) An example of BFS Algorithm; (b) An example of DFS Algorithm

Running previous mentioned algorithms (DFS and BFS) will obtain all possible trees for each node. The next step, is finding out induced sub-graphs for nodes of each tree. After that, inasmuch as the nodes of the tree and the induced sub-graph are completely same, there is at least a cycle between these nodes if the edges in an induced sub-graph are not exactly same as those edges involved in the tree. Since suspected drivers do not require to organize staged collisions in right order (sequence), there is no cycle (n>3) for directed network, it is verified after running algorithms for real data. Note that, the designed algorithms check whether the detected cycles are repetitive or not. Indeed, final cycles will be unique. Besides, short cycles (n = 3) should not be considered because the undirected network of accidents includes many structures such these cycles, while they did not occur with the previous purpose. Since we are looking for organized groups (more sophisticated structures) causing more the revenue leakage and financial losses, these cycles have no worth for us.

Alternatively stated, in the undirected network of car accidents, there is a significant number of cycles with just three nodes, whereas their occurrences are quite normally with the least likelihood of fraudulent activities. As a result, cycles with the number of vertices more than three (n>3) should be considered. In large networks with high volume of data, detections of such components are time-consuming and costly, whereas it does not make it easier for insurance experts. Furthermore, the large cycles (n>50) are also meaningless in detection of organized groups because the occurrence of such cycles seems accidentally (not by arrange). Some example of cycles can be seen in figure 6.

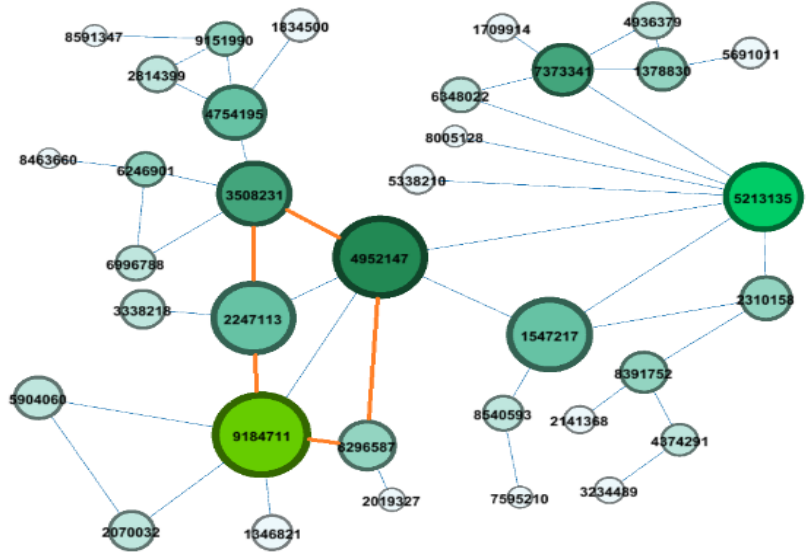

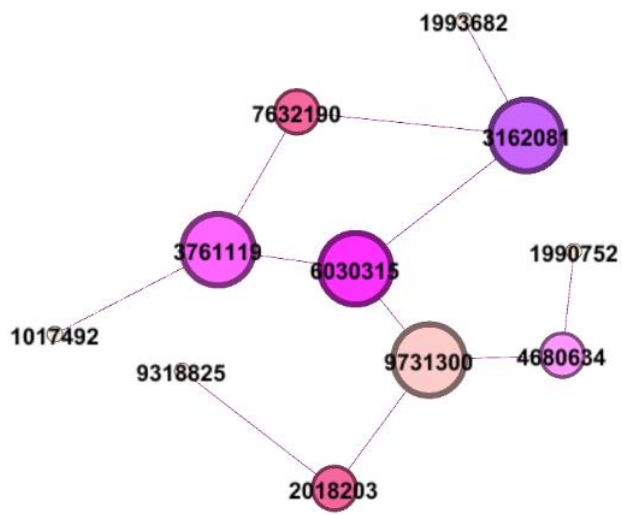

**Figure 6.** Suspicious component detected (having cycles)

*4-3-Assessment and visualization of suspicious entities*

In the last module, the identified cycles (suspected components in previous module) should be investigated much more precisely so that groups with the high likelihood of fraud become revealed. The purpose of the final step is to reduce the required time and later actions of experts for checking all groups of drivers in terms of fraud. According to studies, we use some indicators by which the likelihood of being fraudulent for suspicious components are specified, thus, the maximum likelihood could be considered. The reasons for choosing indicators have been touched on in paper of (Šubelj et al. 2011). All detected cycles ($3<n<50$) could be assessed using a set of indicators defined by experts.

Let $W = [W_1, W_2, ..., W_n]$ ($i = 1, ..., n$) is an ensemble of indicators, and c be some identified cycles in previous section that $c \in C(G)$ (a set of all cycles in graph G). Now, consider $I_i(c)$ as the value for c which is measured by indicator $W_i$, so:

$$W_i(c) = \begin{cases} 1 & \text{if c has suspicious value of } I_i(c), \\ 0 & \text{otherwise} \end{cases} \quad (1)$$

then,

$$P_i(c) = \sum_{i=1}^{n} W_i(c) \quad (2)$$

When all of the most highly likelihood cycles (groups of fraudsters) become flagged, they are visualized, and the investigator or experts analyze these groups and later actions was determined for finding satisfactory ways of addressing potential fraud.

## 5- Experimental Results

The system was implemented on a real data for evaluating the performance of proposed method. In this case study, the applied data includes the data set a period of five years (2007-2012) by a real insurance company in Iran. In the research, police officers, lawyers, insurance workers, participants, garage mechanics, chiropractors, and others were ignored, while only drivers have been taken into identification. This information includes 1,850,928 collisions that consist of 2,308,571 drivers. Drivers only have been considered as nodes of the undirected network of collisions, and their presence in a collisions are shown as edges.

As the beginning of the method, the undirected network corresponding to collisions was generated that the information of network before and after the removal of nodes (d(v) = 0 or 1) is presented in table 1.

**Table 1.** The information about network of collisions for drivers

|  | Numbers of nodes | Numbers of edges | Minimum degree | Maximum degree |
|---|---|---|---|---|
| Total network | 2,308,571 | 1,943,559 | 0 | 49 |
| After removing the nodes ( n = 0 or 1) | 992,075 | 1,360,070 | 2 | 46 |

In order to reveal cycles two different algorithms using DFS and BFS tree have been applied. The data set consists of many cycles that some examples of them can be seen in figure 7, and table 2 shows the results of running algorithms.

**Table 2.** Performance of cycle detection algorithms on network of collisions

|  | DFS | BFS |
|---|---|---|
| Running time n>5 | 5:35:41 | 6:32.05 |
| Total Numbers of rings | 35703 | 35703 |
| 5<n<10 | 23118 | 23118 |
| n>=10&N<50 | 12105 | 12105 |
| N>=50 | 480 | 480 |
| N>=100 | 123 | 123 |
| N>=150 | 57 | 57 |
| N>=200 | 38 | 38 |
| M>=500 | 6 | 6 |

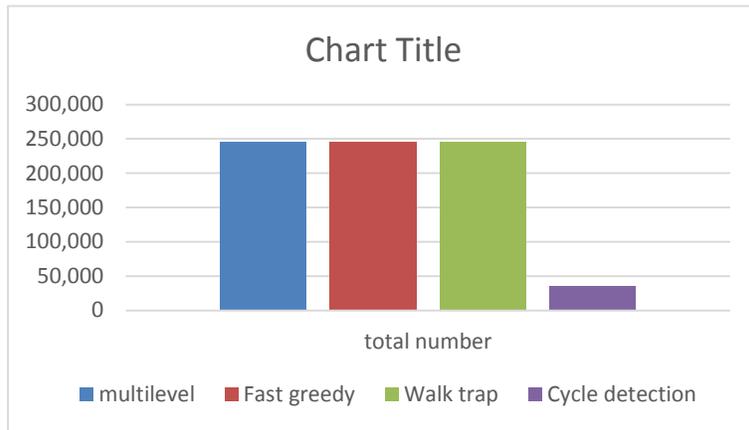

**Figure 7.** Comparison of cycle detection approach against some well-known community algorithms.

According to table 2, the number of detected cycles through two algorithms are same, while the running times are different. Although there are many methods for finding groups (communities) in a network out, the detection of cycles is much more effective because a small number of entities would be considered, while the likelihood of fraud is more for them. Three known community detection algorithms were implemented to prove our claim that detection of cycles (n>3) closes experts upon identifying organized groups of fraudsters. We considered communities detected through the multilevel, fast greedy, and walk trap (see table 3, and figure 8).

**Table 3.** Comparison of different approach for detection of fraudulent groups against detection of cycles

| algorithms | multilevel | Fast greedy | Walk trap | cycle |
|---|---|---|---|---|
| Total number | 245,713 | 245,714 | 245,709 | 35703 |

A glance at the provided table and bar chart demonstrates that there are huge differences between the numbers of discerned cycles compared to discovered communities. Clearly, these differences make assessment too much easy for investigators and experts regarding groups of perpetrators because they could contemplate a small number of entities causing investigation that is more precise and less time consuming. Next, we analyzed several suspicious cycles by the same set of six indicators was applied for calculation the probability of fraud.

**Table 4.** Fraudulent likelihood of some detected cycles

| Cycle ID | nodes | edges | density | Fraud Probability |
|---|---|---|---|---|
| 5136 | 19 | 23 | 0.135 | 6.798 |
| 1385 | 12 | 19 | 0.288 | 6.506 |
| 5072 | 15 | 15 | 0.143 | 5.864 |
| 7428 | 18 | 18 | 0.118 | 5.809 |
| 3748 | 13 | 12 | 0.154 | 5.781 |
| 3891 | 17 | 16 | 0.118 | 5.486 |
| 1952 | 15 | 15 | 0.143 | 5.332 |
| 4721 | 10 | 9 | 0.200 | 5.306 |
| 3206 | 12 | 12 | 0.182 | 4.241 |
| 6790 | 8 | 10 | 0.357 | 5.141 |
| 4508 | 13 | 14 | 0.180 | 5.115 |
| 1487 | 20 | 20 | 0.105 | 8.015 |
| 46680 | 16 | 19 | 0.158 | 7.804 |
| **61656** | 20 | 19 | 0.100 | 7.768 |
| **61511** | 19 | 18 | 0.105 | 7.456 |
| **33906** | 18 | 20 | 0.131 | 7.353 |
| **75538** | 15 | 15 | 0.143 | 6.832 |
| **73428** | 16 | 16 | 0.133 | 5.540 |
| **70541** | 20 | 20 | 0.105 | 5.450 |
| **37441** | 11 | 10 | 0.182 | 6.403 |
| **38019** | 16 | 18 | 0.150 | 6.383 |
| **36148** | 12 | 11 | 0.167 | 6.269 |

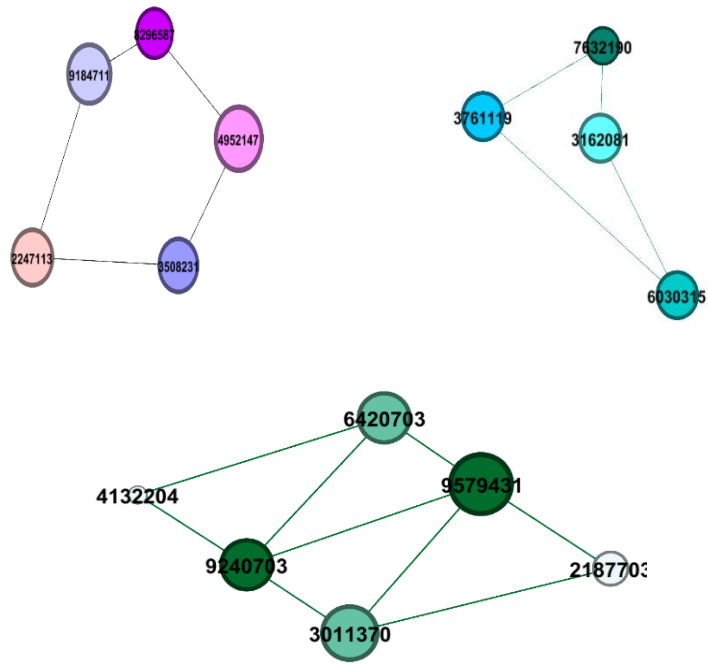

**Figure 8.** Detected group with high likelihood

The table 4 and figure 8 demonstrate the number of components with high likelihood. Note that, the fraud likelihood for suspicious components whose number of their nodes is more than (50), were highly low shown the occurrences of them are coincidentally. Then, we compared the likelihood of fraud for several founded communities using mentioned algorithms previously including cycles with the fraud likelihood of their cycles (see table 5, figure 7). Overall, according to comparison results, the likelihoods of fraud for suspicious cycles are higher in comparison with communities in which the cycles were detected.

**Table 5.** Comparison of Fraud likelihood for several communities in which detected

| Community ID | Number of nodes | Fraud probability | Cycle ID | Number of nodes | Fraud probability |
|---|---|---|---|---|---|
| **51814** | 23 | 2.898 | 1573 | 10 | 7.898 |
| **74230** | 34 | 2.605 | 6314 | 14 | 6.905 |
| **14496** | 19 | 1.868 | 2855 | 5 | 4.868 |
| **83761** | 13 | 1.812 | 8426 | 8 | 3.812 |
| **94355** | 10 | 3.759 | 7051 | 6 | 2.759 |
| **25065** | 25 | 0.638 | 6370 | 12 | 1.638 |

**6-Summary and Conclusion**

In this paper, we proposed a method for detection of organized groups of perpetrators in automobile insurance. Although many methods have been put forward so far with respect to flagging such as groups, none of them is effective enough for a large size data set. Since the social network analysis is a suitable method for modeling the relationships between entities, it has been applied for revealing professional frauds. Moreover, cycles play outstanding roles in networks' structures, thus, we technically considered them. Doubtless the likelihood of being fraudulent for drivers whose relations form in a network of collisions is as a cycle, are more than other entities. Our proposed method is to generate the undirected network of collisions at beginning stage. Afterward, distinguishing suspicious cycles that the number of their nodes is more than three (3<n) because the occurrence of such cycles might be with done planning previously and bring more revenue leakage for insurance companies. Note that the cycles whose number of nodes are either equal to 3 or more than 50 were ignored inasmuch as their artificial occurrence in undirected networks. The following stage is to investigate suspicious cycles with computing the likelihood of fraud through introducing several indicators in order to prioritize later review for experts.

Finally, fraudulent groups were detected and visualized with the least time, and the volume of calculation for experts saw a remarkable decrease. To evaluate the proposed method, we compared the achieved result with output from other systems comprising some community detection algorithms. In the near future, we plan to extend our method to other fields like credit card or telecommunications.